\documentclass[prd,nofootinbib]{revtex4}

\usepackage{amsmath,amssymb}
\usepackage{mathrsfs}
\usepackage{pstricks}
\usepackage{pst-func}
\usepackage{graphicx}
\usepackage{refcount}
\usepackage{cases}

\def\dbl{\hbox{${1\hskip -2.4pt{\rm l}}$}}

\allowdisplaybreaks

\begin{document}

\title{Local Quantum Mechanical Prediction of the Singlet State Using Geometric Algebra}

\author{Carl F. Diether, III}

\email{fred.diether@einstein-physics.org}

\affiliation{Einstein Centre for Local-Realistic Physics, Santa Rosa, CA, USA}

\date{11-18-2025}

\begin{abstract}
We deduce the quantum mechanical prediction of $-{\bf a}\cdot{\bf b}$ for the singlet spin state employing local measurement functions following Bell's approach.  This result represents the quantum mechanical expectation value for the joint measurement of spin projections in the singlet state. And is equal to the negative cosine of the angle between vectors {\bf a} and {\bf b}. Our derivation is corroborated through a computational simulation conducted in the Mathematica programming environment using geometric algebra. 
\end{abstract}

\maketitle

\parskip 6pt
\baselineskip 12pt

\section{Introduction}
In this manuscript, we elucidate the quantum mechanical (QM) prediction for the singlet spin state in the context of the EPR-Bohm thought experiment (EPRB) \cite{Bohm1957, EPR1935}, employing local measurement functions in accordance with the methodologies of Bell \cite{Bell-1964} and Christian \cite{Christian2018}. Within the framework of the EPRB experiment, we consider two spin one-half quantum particles (an electron and a positron) propagating in opposite directions subsequent to their generation from a singlet state particle. Subsequently, at Stations A and B situated at a spacelike separation measurements of their spins are conducted along independently chosen unit vectors ${\bf a}$ and ${\bf b}$. We posit that the progenitor particle of the pair possesses zero spin, thereby endowing the pair with the following singlet wavefunction \cite{GHSZ}:
\begin{equation}
|\Psi_{\bf n}\rangle
=\frac{1}{\sqrt{2}}\Bigl\{|{\bf n},\,+\rangle_1\otimes
|{\bf n},\,-\rangle_2\,-\,|{\bf n},\,-\rangle_1\otimes|{\bf n},\,+\rangle_2\Bigr\}, 
\end{equation}
where
\begin{equation}
{\boldsymbol\sigma}\cdot{\bf n}\,|{\bf n},\,\pm\rangle\,=\,
\pm\,|{\bf n},\,\pm\rangle, \label{spinA}
\end{equation}
characterizes the quantum mechanical eigenstates where the particles exhibit spin in the "up" or "down" orientation measured in units of ${\hbar=2}$. Here, ${\boldsymbol\sigma}$ denotes the Pauli spin vector ${({\sigma_x},\,{\sigma_y},\,{\sigma_z})}$, while ${\bf n}$ represents the singlet particle's spin vector oriented in an arbitrary 3-dimensional direction.  And ${\bf n}$ is transferred to the two particles created from the singlet particle. 

The rotational symmetry inherent in the singlet state $\Psi_{\bf n}$ guarantees that the expectation values for the individual spin observables ${{\boldsymbol\sigma}\cdot{\bf a}}$ and ${{\boldsymbol\sigma}\cdot{\bf b}}$ are
\begin{align}
{\cal E}_{q.m.}({\bf a})\,&=\,\langle\Psi_{\bf n}|\,{\boldsymbol\sigma}
\cdot{\bf a}\otimes\dbl\,|\Psi_{\bf n}\rangle \notag \\
&=\,\langle\Psi_{\bf n}|\,{\boldsymbol\sigma}
\cdot{\bf a}\,|\Psi_{\bf n}\rangle\,=\,0 \\
\text{and}\;\;\;{\cal E}_{q.m.}({\bf b})\,&=\,\langle\Psi_{\bf n}|\,\dbl\otimes{\boldsymbol\sigma}
\cdot{\bf b}\,|\Psi_{\bf n}\rangle \notag \\
&=\,\langle\Psi_{\bf n}|\,{\boldsymbol\sigma}
\cdot{\bf b}\,|\Psi_{\bf n}\rangle\,=\,0\,,\label{rotinvar}
\end{align}
where ${\dbl}$ is the identity matrix.  The expectation value of the joint observable
${{\boldsymbol\sigma}\cdot{\bf a}\otimes{\boldsymbol\sigma}\cdot{\bf b}}$ is \cite{GHSZ,Yablon2019-2}
\begin{equation}
{\cal E}_{q.m.}({\bf a},\,{\bf b})\,=\,
\langle\Psi_{\bf n}|\,{\boldsymbol\sigma}\cdot{\bf a}\,\otimes\,
{\boldsymbol\sigma}\cdot{\bf b}\,|\Psi_{\bf n}\rangle\,=\,-\,
{\bf a}\cdot{\bf b}\,,\label{twoobserveA1}
\end{equation}
regardless of the relative distance between the two remote locations represented by the unit vectors ${\bf a}$ and ${\bf b}$.  We will match this result with a calculation of the product expectation value of the local measurement functions.  Nevertheless, it is posited that the state, $\Psi_n$, used by this equation ceases to exist upon and after detection, raising questions regarding the validity of this as an actual physical process within the framework of quantum mechanics.

We examine an additional derivation found in Appendix B of the GHSZ paper \cite{GHSZ}. This corresponds to their eq. (B3) transposed into contemporary notation.
\begin{align}
|\Psi\rangle = \frac{1}{\sqrt{2}} &\left[ -\left(\sin \frac{\theta}{2}\right) |{\bf n_1}, +\rangle_1 |{\bf n_2}, +\rangle_2 \right.
 + \left(\cos \frac{\theta}{2}\right) |{\bf n_1}, +\rangle_1 |{\bf n_2}, -\rangle_2 \notag \\ 
&\;\, - \left(\cos \frac{\theta}{2}\right)  |{\bf n_1}, -\rangle_1 |{\bf n_2}, +\rangle_2  
 \left. - \left(\sin \frac{\theta}{2}\right)  |{\bf n_1}, -\rangle_1 |{\bf n_2}, -\rangle_2  \right].
\end{align}
Where $\theta$ is the polar angle between ${\bf n_1}$ and ${\bf n_2}$.  From that equation, they deduce the joint outcome probabilities.
\begin{align}
P^{\psi}_{++}({\mathbf{n}}_1,{\mathbf{n}}_2) &= \frac{1}{2}\sin^2 \theta/2, \\
P^{\psi}_{+-}({\mathbf{n}}_1,{\mathbf{n}}_2) &= \frac{1}{2}\cos^2 \theta/2, \\
P^{\psi}_{-+}({\mathbf{n}}_1,{\mathbf{n}}_2) &= \frac{1}{2}\cos^2 \theta/2, \\
P^{\psi}_{--}({\mathbf{n}}_1,{\mathbf{n}}_2) &= \frac{1}{2}\sin^2 \theta/2.  
\end{align}
Then sum them together for the expectation value result according to their eq. (2),
\begin{equation}
    E^{\Psi}({\mathbf{n}}_1,{\mathbf{n}}_2) = \sin^2 \theta/2 - \cos^2 \theta/2 = -\cos \theta = -\mathbf{n}_1 \cdot \mathbf{n}_2.
\end{equation}
This appears paradoxical since ${\bf n_1}$ and ${\bf n_2}$ must correspond to our ${\bf a}$ and ${\bf b}$ respectively. This raises a question regarding the persistence of such a correspondence when the singlet state no longer prevails at the same instance.  This situation is similar to eq. (5) but both obtain the correct result according to our analysis.  We will discuss this in more detail in the conclusion. 

In 1964, John S. Bell formulated a precise test to distinguish local hidden-variable theories from quantum mechanics. Bell showed that any theory in which measurement outcomes are determined by pre-existing (“hidden”) parameters, and in which no influence travels faster than light, must satisfy certain statistical bounds—now known as Bell inequalities \cite{Bell-1964}. These inequalities constrain the strength of correlations between distant measurement outcomes, regardless of the detailed mechanism of the hidden variables. Quantum mechanics, by contrast, predicts and experiments confirm violations of these inequalities through correlations such as $\langle A({\bf a})B({\bf b})\rangle = - {\bf a}\cdot {\bf b}.$  
Bell’s result thus established that either locality or realism (or both) must be abandoned if one accepts the empirical success of quantum theory. His framework also introduced the notion of explicit measurement functions, $A({\bf a}, \lambda)$ and $B({\bf b}, \lambda)$, laying the groundwork for any local realistic model to be tested against quantum predictions.  But as we shall see, it is not necessary to abandon locality for quantum mechanics.

Building on Bell’s measurement function formalism, Christian has proposed a local realistic reconstruction of singlet state correlations using the algebra of quaternions or geometric algebra underpinning the topology of the 3-sphere \cite{Christian2018, Christian2014}. In Christian’s model, each particle carries a hidden quaternionic “spinor” $\lambda$ living on $S^3$ with orientation, which serves as an analogue of a hidden variable in Bell’s formalism. The detection process is represented by mappings from $S^3$ to $\pm 1$ that depend only on the local detector setting and $\lambda$. The non-trivial 3-sphere geometry endows these maps with just the right cross- and dot-product structure to reproduce the quantum result $- {\bf a}\cdot {\bf b}$, while strictly respecting locality. Christian’s approach leverages Geometric Algebra to unify scalars, vectors, and bivectors in a single formalism, showing that the standard quantum correlation emerges naturally from the underlying quaternionic topology of the 3-sphere.  For the physical significance of the 3-sphere topology and quaternions used in our measurement functions in much detail, see the references above and \cite{Christian2, Christian1, Christian3, Christian4}.

\section{Formulation of Measurement Functions}

We shall now formulate certain explicit local measurement functions, similar to those in Bell's framework \cite{Bell-1964}. This approach yields the aforementioned result of $-{\bf a}\cdot{\bf b}$ and aligns with the eigenvalues of observable operators that entail spins being detected by detectors, with the division of the singlet vector governed by the conservation of spin angular momentum, as discussed with ${\bf s_1}+{\bf s_2}=0$, so that:
\begin{equation}
{\bf n} \,= {\bf s} \,= {\bf s}_1 = -{\bf s}_2. \label{eq6}
\end{equation}
Upon detection, ${\bf s_1}$ and ${\bf s_2}$ become $ {\bf a}$ and $ {\bf b}$, respectively.

In the interaction of two vectors, the resultant is composed of a scalar, $({\mathbf u}\cdot {\mathbf v})$, in addition to a vector via a cross-product, $({\mathbf u}\times {\mathbf v})={\mathbf r}$, which may collectively be represented as a quaternion, ${\mathbf q}({\mathbf u}\cdot {\mathbf v},\,{\mathbf r})$, as a single entity.  This can also be done in geometric algebra with multi-vectors $({\bf u}\,\;{\bf v}) = ({\bf u}\cdot {\bf v}+{\bf u}\wedge {\bf v})$ \cite{Doran}.  Employing this notion in conjunction with eq.(\ref{eq6}), we define:
\begin{align}
  {\bf r_a} &\,:= ({\bf a}\times {\bf s}_1) \;\;\text {and}\;\;  {\bf r_b} := ({\bf s}_2 \times {\bf b}), \\
  \mu_{\bf a} &:= \text{sgn}({\bf a}\cdot {\bf s}_1){\bf a}, \\
  \mu_{\bf b} &:= \text{sgn}({\bf b}\cdot {\bf s}_2){\bf b}, \\
  |\phi_{\bf s_1}\rangle
  &:=\frac{1}{\sqrt{2}}\Bigl\{|{\bf s_1},\,+\rangle_1 + |{\bf s_1},\,-\rangle_1\Bigr\}, \\
  |\chi_{\bf s_2}\rangle
  &:=\frac{1}{\sqrt{2}}\Bigl\{|{\bf s_2},\,+\rangle_2 + |{\bf s_2},\,-\rangle_2\Bigr\},\\
  A({\bf a},\, {\bf s_1}) &:= \lim_{{\bf s_1} \rightarrow \, \mu_a}\Big[\langle\phi_{\bf s_1}|({\boldsymbol\sigma}\cdot {\bf a}) ({\boldsymbol\sigma}\cdot {\bf s_1})|\phi_{\bf s_1}\rangle + I_3\,{\bf r}_a\, \Big], \notag \\
  &\,=\, \lim_{{\bf s_1} \rightarrow \, \mu_a}\Big[{\bf a}\cdot{\bf s_1} + I_3\,{\bf r}_a\, \Big], \notag \\
  &\,=\, \lim_{{\bf s_1} \rightarrow \, \mu_a}\Big[{\bf a}\,\;{\bf s_1} \Big], \notag \\
  &\,=\, \text{sgn}({\bf a}\cdot{\bf s_1})= \pm 1, \\
 B({\bf b},\, {\bf s_2}) 
  &:=\lim_{{\bf s_2} \rightarrow \, \mu_b}\Big[\langle\chi_{\bf s_2}| ({\boldsymbol\sigma}\cdot {\bf s_2}) ( {\boldsymbol\sigma}\cdot {\bf b})|\chi_{\bf s_2}\rangle + I_3\,{\bf r}_b\,  \Big], \notag \\ 
  &\,=\, \lim_{{\bf s_2} \rightarrow \, \mu_b}\Big[{\bf s_2}\cdot{\bf b} + I_3\,{\bf r}_b\,  \Big], \notag \\
  &\,=\, \lim_{{\bf s_2} \rightarrow \, \mu_b}\Big[{\bf s_2}\,\;{\bf b} \Big] ,\notag \\
  &\,=\, \text{sgn}({\bf s_2}\cdot{\bf b})= \mp 1,
\end{align}

And where $I_3 = {\bf e_x e_y e_z}$, which is the pseudo-scalar of Geometric Algebra \cite{Doran}.  The derivation of the expectation value of the calculation pertaining to the measurement functions A and B is detailed in {\bf Appendix A}. In this context, ${\boldsymbol\sigma}\cdot{\bf a}$ and ${\boldsymbol\sigma}\cdot{\bf b}$ are designated as the detectors utilized by Alice and Bob, respectively, each possessing no angular momentum at the instant of detection. ${\boldsymbol\sigma}\cdot{\bf s_1}=-{\boldsymbol\sigma}\cdot{\bf s_2}$ denotes the spin of the fermions these detectors receive, which constitutes the basis for conducting the EPRB experiment. The limit replacement functions simulate the functioning of the polarizers at the detection stations and are clearly notation simplification, while $|\phi_{\bf n}\rangle$ and $|\chi_{\bf n}\rangle$ are now the wavefunctions of the individual particles. In the subsequent stage of the A and B functions, it becomes apparent that multi-vectors are delineated from the preceding stage. The cross-products vanish in the matrix expectation value calculation due to averaging.  And they should not vanish according to vector algebra therefore we assume they are not observable but need to be included in the overall calculation for the prediction as can be seen in {\bf Appendix B}.

\section{Product Expectation Value Calculation}

The measurement functions delineate the detection processes occurring at two potentially spacelike-separated observation stations belonging to Alice and Bob. Despite possibly transpiring concurrently, $A({\bf a},\, {\bf s_1})$ and $B({\bf b},\,{\bf s_2})$ constitute independent physical processes but are subject to the conservation of the initial zero spin angular momentum from the source particle. In other words, if ${\bf a = b}$ then the functions will be opposite each other. In advancing the $AB$ product calculation, the $k$ indices are omitted subsequent to the initial step. Employing the "product of limits equal to limits of product" principle results in the computation of the expectation value as follows \cite{Christian1}:

\begin{align}
{\cal E}({\bf a},\,{\bf b})\,&=\lim_{\,n>>1}\left[\frac{1}{n}\sum_{k\,=\,1}^{n}\,{A}({\bf a}^k,\,{\bf s}_1^k)\;{B}({\bf b}^k,\,{\bf s}_2^k)\right], \\
&=\!\!\lim_{\,n\,\gg\,1}\Bigg\{\frac{1}{n}\sum_{k\,=\,1}^{n}\left[\lim_{{\mathbf s}_1\,\rightarrow\,\mu_a}\left\{{\bf a}\,\;{\bf s_1}\right\}\right]
\left[\lim_{{\mathbf s}_2\,\rightarrow\,\mu_b}\left\{{\bf s_2}\,\;{\bf b}\right\}\right]\!\Bigg\}, \label{10-2n} \\
&=\!\!\lim_{\,n\,\gg\,1}\Bigg[\frac{1}{n}\sum_{k\,=\,1}^{n}\;\lim_{\substack{{\mathbf s}_1\,\rightarrow\,\mu_a \\ {\mathbf s}_2\,\rightarrow\,\mu_b}}\Big\{({\bf a}\,\;{\bf s_1})\;({\bf s_2}\,\;{\bf b})\Big\}\Bigg], \label{11-2n} \\
&=\!\!\lim_{\,n\,\gg\,1}\Bigg[\frac{1}{n}\sum_{k\,=\,1}^{n}\;\lim_{\substack{{\mathbf s}_1\,\rightarrow\,\mu_a \\ {\mathbf s}_2\,\rightarrow\,\mu_b}}\Big\{({\bf a}\cdot {\bf s_1}+I_3 {\bf r_a})\;({\bf s_2}\cdot {\bf b} + I_3 {\bf r_b})\Big\}\Bigg], \label{11-4n} \\
&=\!\!\lim_{\,n\,\gg\,1}\Bigg[\frac{1}{n}\sum_{k\,=\,1}^{n}\;\lim_{\substack{{\mathbf s}_1\,\rightarrow\,\mu_a \\ {\mathbf s}_2\,\rightarrow\,\mu_b}}
\Big\{-{\bf a}\cdot {\bf b}+{\bf r_0} \Big\}\Bigg], \label{11-3n}\\
&=\!-{\mathbf a}\cdot {\mathbf b}+\!\lim_{\,n\,\gg\,1}\Bigg[\frac{1}{n}\sum_{k\,=\,1}^{n}\;\lim_{\substack{{\mathbf s}_1\,\rightarrow\,\mu_a \\ {\mathbf s}_2\,\rightarrow\,\mu_b}}
\Big\{{\mathbf r}_0 \Big\}\Bigg], \label{78-n}\\
&=-{\mathbf a}\cdot {\mathbf b} + \vec{\,\mathbf 0},\label{18}  \\
&= -{\mathbf a}\cdot {\mathbf b}, \label{anacor} \\
\text{where}\quad {\bf r}_0\,({\bf s}_1,\, {\bf s}_2) &= I_3\{({\bf a}\cdot {\bf s}_1)({\bf s}_2 \times {\bf b})+({\bf s}_2 \cdot {\bf b})({\bf a} \times {\bf s}_1)-(({\bf a} \times {\bf s}_1) \times ({\bf s}_2 \times {\bf b}))\},
\end{align} 
In step eq.(\ref{11-3n}), we have employed an identity pertinent to the multiplication of multi-vectors, as referenced in \cite{Christian2,Christian1,Christian3}. The proof of the identity and of eq.(\ref{11-3n}) can be found in {\bf Appendix B}. It is noted that ${\bf r}_0$ contains all cross-products that become zero upon taking the limits. Thus in step (\ref{78-n}), the limit replacement functions are applied exclusively to the cross-products of ${\bf r}_0$, resulting in a null vector. Consequently, the desired result is obtained through a wholly local process. We have verified the above analytical calculation through a computer simulation using Mathematica programming language with geometric algebra \cite{Diether} and the result can be seen in Figure 1. The Mathematica notebook file is available to read or download at \cite{Diether}.

\section{Computational Validation via Mathematica Simulation}

The Mathematica simulation explores the validation of the local quantum mechanical analytical prediction of the singlet state using advanced mathematical tools such as Pauli matrices, Clifford Algebra, and 3D vectors, inspired by Dr. Christian's 3-Sphere Model \cite{Christian2018}. It includes Mathematica simulations for generating particle spin data, analyzing particle-detector interactions, and verifying predictions through geometric algebra calculations. The correlation data from the simulation are compared with a negative cosine curve ($-{\bf a}\cdot {\bf b}$) for validation and is an exact match. The results compute averages and discuss cross-product behaviors in relation to predictions.  An in-depth examination of the principal components is presented below \cite{Diether}:

\noindent \textbf{A. Simulation Framework}\\
The simulation follows the setup of an EPR-Bohm experiment, modeling two spin-1/2 particles emitted in opposite directions from a singlet state particle. This computational approach challenges conventional quantum interpretations by demonstrating that the expected correlation, $-{\bf a}\cdot {\bf b}$, can emerge from a purely local framework.

\noindent To verify the predicted expectation value, the simulation:\\
1. Generates random spin vectors for two entangled particles, ensuring uniform distribution over the unit sphere.\\
2. Implements Pauli matrix operations to define measurement interactions.\\
3. Converts spin interactions into quaternion representations for computation.\\
4. Evaluates measurement functions $A({\bf a}, {\bf s_1})$ and $B({\bf b}, {\bf s_2})$ to derive correlation values.\\
5. Compares the numerical results with the expected theoretical function.

\noindent \textbf{B. Implementation Details}\\
{1. Particle Spin Initialization}  \\
   - The simulation generates 30,000 random spin vectors, with each spin pair constrained by the singlet state condition:  
     $s_2 = -s_1$  \\
   - Each spin interacts with measurement devices, encoded using Pauli matrices.\\
2. Measurement Interaction and Geometric Algebra Representation  \\
  - The particle-detector interaction is computed using:\\       
  $A({\bf a, s_1}) = \text{sgn}({\bf a} \cdot {\bf s_1})$ and
     $B({\bf b, s_2}) = \text{sgn}({\bf b} \cdot {\bf s_2})$ so that averages can be performed.\\
   - Measurement vectors are transformed via geometric algebra, allowing efficient manipulation of spin orientations.\\
3. Calculation of the correlation function \\
- The expectation value ${\cal E}({\bf a,\, b})$ is computed on the basis of the inner products and cross-products of the spin vectors.  
   - The simulation verifies that the final correlation function matches the negative cosine curve, confirming agreement with standard QM predictions.

\noindent \textbf{C. Steps and Results in the Simulation}\\
1. Spin Data Generation:\\
   - Using random points on a unit sphere, the spin vectors of particles directed to detectors A and B are simulated. This ensures uniformity in vector distribution.\\
2. Detector Interactions:\\
   - The interaction between spin vectors and detectors involves complex quaternion calculations. \\
3. Correlation Analysis:\\
   - The simulation compares the calculated data with a negative cosine curve to validate the local QM predictions. A visual plot includes:
     - Blue: Simulated correlation data.\\
     - Magenta: Negative cosine function.\\
4. Averages and Cross Products:\\
   - Individual outcomes for the A and B measurement functions average to zero, aligning with QM predictions. \\
     - ⟨A⟩ = 0.00227\\
     - ⟨B⟩ = 0.00767\\
     - Cross products vanish by the limit functions: mean(pc) = -0.000965.

\noindent \textbf{D. Significance of the Results}\\
This simulation validates that a purely local quantum mechanical prediction reproduces the expected singlet state correlation, demonstrating alignment with standard quantum theory without requiring nonlocal influences. The results strengthen the analytical derivation, confirming that local measurement functions yield correct quantum correlations.

Figure 1 illustrates the correlation data of the simulation, precisely overlaying the theoretical negative cosine curve, reinforcing the correctness of the local QM approach.
The results highlight that the simulation is consistent with our local quantum-mechanical analytical prediction and demonstrates compatibility with the theoretical framework.

\begin{figure}[h]
\vspace{0.2cm}
\centering
\includegraphics[scale=0.6]{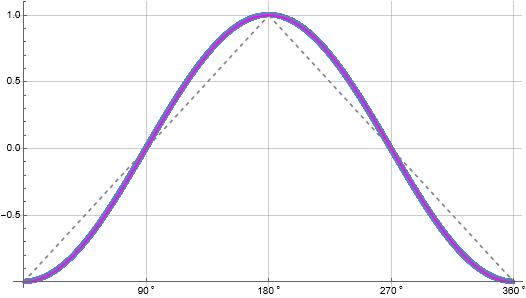}
\caption{Plot of product calculation from the simulation.  Blue is the  correlation data, magenta is the negative cosine curve for an exact match.}
\vspace{0.2cm}
\label{Fig-1}
\end{figure}

\section{Conclusion}

Certain individuals assert that our prediction should align with experimental outcomes on an individual occurrence basis. However, this methodology does not align with the predictive framework of quantum mechanics. Quantum mechanics articulates its predictions regarding $-{\bf a}\cdot{\bf b}$ for EPRB through the use of statistical averages, either as probability averages or expectation values \cite{GHSZ}. In accordance with this approach, we also formulate our prediction based on averages.

Locality in quantum mechanics first became a central concern with the 1935 paper by Einstein, Podolsky and Rosen \cite{EPR1935}.  In their EPR paradox, they showed that quantum mechanics predicts perfect correlations between distant particles, suggesting “spooky action at a distance”—i.e. one measurement instantaneously affecting the other—seemingly in conflict with the relativistic prohibition on faster-than-light influences.  The issue was sharpened in 1964 when John S. Bell derived his famous inequalities \cite{Bell-1964}.  Bell proved that any theory based on local hidden variables must satisfy certain statistical bounds—bounds that quantum mechanics violates through correlations like $\langle A({\bf a})B({\bf b})\rangle = - {\bf a}\cdot {\bf b}.$  This result put locality on firm experimental footing by showing that quantum predictions (and experiments) are incompatible with any strictly local realistic theory.  However, Christian disproved Bell's theorem in 2007, which was not accepted by mainstream physics, but the result was eventually published in \cite{Christian2018, Christian4}, two mainstream journals.

As I have demonstrated, it is feasible to arrive at a solution for a local QM result. The solution is achieved through the application of 3-sphere topology using 3D vectors, geometric algebra, and the conservation of angular momentum. This discourse addresses the presumptions made by John Bell regarding the non-local nature of quantum mechanics in the EPRB scenario, as can be seen in \cite{Bell-2004}. Nonetheless, Bell provided valuable insights into the formulation of local measurement functions. Subsequently, Joy Christian refined the formulation by incorporating limit replacement functions, which facilitated accurate product calculations \cite{Christian2018,Christian2014,Christian2, Christian1}. It is probable that previous assertions claiming ``ordinary quantum mechanics is not locally causal" \cite{Bell-2004} may be incorrect at least for the EPRB thought experiment.

Although both Equation (5) and the derivation in {\bf Appendix B} of the GHSZ manuscript successfully reproduce the prediction of $-{\bf a}\cdot {\bf b}$, they highlight a profound conceptual tension. The coincidence of the results, despite the seeming non-equivalence of the initial states---the enduring singlet state in one case versus the \textit{a priori} pre-determined state in the other---suggests a deeper physical reality at play.  Both omit the stage where the alignment of spin vectors with the polarizers of the detectors occurs.  The problem is a timing issue since the vectors ${\bf a}$ and ${\bf b}$ do not exist at the same time as the singlet state wavefunction.  Therefore, it is posited that our methodological approach yields a process that aligns more closely with physical reality.

The core of the matter lies in the post-detection condition. As posited, the state of the system is fundamentally altered upon measurement. The mathematical description of the entangled singlet state, while powerful for predicting pre-measurement probabilities, ceases to apply at the moment of pair detection. This raises critical questions about the nature of the correspondence between the initial measurement settings (${\bf a}, {\bf b}$) and the properties of the detected particles (${\bf s_1}, {\bf s_2}$). The paradox arises because our mathematical formalism, which assumes a persistent state, yields the correct expectation values, while a physical interpretation demands the state's collapse.

The agreement between the two derivations, therefore, provides a strong basis for questioning the persistence of the singlet state after detection. The consistency in the results serves not as a validation of an enduring state but rather as an indicator that the underlying physical process is more complex. It suggests that the correlation is not maintained by a continuous, nonlocal connection but is an emergent property that manifests precisely when the state is no longer a superposition. Ultimately, this analysis underscores the need to re-examine the classical assumptions of preexisting reality in the context of quantum measurement.

\section*{Acknowledgments}

The author expresses gratitude to Joy Christian and Jay Yablon for their substantial contributions to this paper, which were offered through numerous discussions.

\section*{Conflicts of Interest}

The author declares no conflicts of interest regarding the publication of this paper.

\appendix
\section{Matrix Expectation Value Calculation for Functions A and B}

We demonstrate here the mathematical steps for the calculation of the matrix expectation value in the measurement functions that produce the inner product with ${\bf s_1}={\bf s} = -{\bf s_2}$.  This calculation was performed using Mathematica for symbolic mathematics.  The computation for Alice's station,
\begin{align}
{\cal E}({\bf a, s_1})\,&=\langle\phi_{\bf s_1}|({\boldsymbol\sigma}\cdot {\bf a})  ({\boldsymbol\sigma}\cdot {\bf s_1})|\phi_{\bf s_1}\rangle,  \\
&=\,\frac{1}{2}\Bigg\{\begin{pmatrix}1&0\end{pmatrix}\begin{pmatrix}{ a_z}&a_x - i a_y \\a_x + i a_y&-{a_z}\end{pmatrix} \begin{pmatrix} { s_z} &s_x - i s_y \\s_x + i s_y& -s_z\end{pmatrix} \begin{pmatrix}1\\0
\end{pmatrix} \notag \\
&\quad\, +\begin{pmatrix}0&1\end{pmatrix}\begin{pmatrix}{ a_z}&a_x - i a_y \\a_x + i a_y&-{a_z}\end{pmatrix} \begin{pmatrix} { s_z} &s_x - i s_y \\s_x + i s_y& -s_z\end{pmatrix} \begin{pmatrix}0\\1
\end{pmatrix}\Bigg\}, \\
&=\,\frac{1}{2}\Bigg\{\begin{pmatrix}1&0\end{pmatrix}\begin{pmatrix}a_x s_x+a_y s_y+a_z s_z +i a_x s_y-i a_y s_x & a_z s_x-a_x s_z-i a_z s_y+ia_y s_z\\-a_z s_x+a_x s_z-i a_z s_y+ia_y s_z&a_x s_x+a_y s_y+a_z s_z -i a_x s_y+i a_y s_x\end{pmatrix} \begin{pmatrix}1\\0
\end{pmatrix} \notag \\
&\quad\, +\begin{pmatrix}0&1\end{pmatrix}\begin{pmatrix}a_x s_x+a_y s_y+a_z s_z +i a_x s_y-i a_y s_x & a_z s_x-a_x s_z-i a_z s_y+ia_y s_z\\-a_z s_x+a_x s_z-i a_z s_y+ia_y s_z&a_x s_x+a_y s_y+a_z s_z -i a_x s_y+i a_y s_x\end{pmatrix} \begin{pmatrix}0\\1
\end{pmatrix}\Bigg\}, \\
&=\frac{1}{2}((a_x s_x +a_y s_y +a_z s_z +i a_x s_y -i a_y s_x)+(a_x s_x +a_y s_y +a_z s_z -i a_x s_y +i a_y s_x)), \\
&= a_x s_x +a_y s_y +a_z s_z = {\bf a}\cdot {\bf s_1}.
\end{align}
Then the computation for Bob's station,
\begin{align}
{\cal E}({\bf s_2, b})\,&=\langle\chi_{\bf s_2}|({\boldsymbol\sigma}\cdot {\bf s_2})  ({\boldsymbol\sigma}\cdot {\bf b})|\chi_{\bf s_2}\rangle,  \\
&=\,\frac{1}{2}\Bigg\{\begin{pmatrix}1&0\end{pmatrix}\begin{pmatrix}{-s_z}&-s_x + i s_y \\-s_x - i s_y&{s_z}\end{pmatrix} \begin{pmatrix} { b_z} &b_x - i b_y \\b_x + i b_y& -b_z\end{pmatrix} \begin{pmatrix}1\\0
\end{pmatrix} \notag \\
&\quad\, +\begin{pmatrix}0&1\end{pmatrix}\begin{pmatrix}{- s_z}&-s_x + i s_y \\-s_x - i s_y&{s_z}\end{pmatrix} \begin{pmatrix} { b_z} &b_x - i b_y \\b_x + i b_y& -b_z\end{pmatrix} \begin{pmatrix}0\\1
\end{pmatrix}\Bigg\}, \\
&=\,\frac{1}{2}\Bigg\{\begin{pmatrix}1&0\end{pmatrix}\begin{pmatrix}-b_x s_x-b_y s_y-b_z s_z -i b_x s_y+i b_y s_x & -b_z s_x+b_x s_z+i b_z s_y-ib_y s_z\\b_z s_x-b_x s_z+i b_z s_y-ib_y s_z&-b_x s_x-b_y s_y-b_z s_z +i b_x s_y-i b_y s_x\end{pmatrix} \begin{pmatrix}1\\0
\end{pmatrix} \notag \\
&\quad\, +\begin{pmatrix}0&1\end{pmatrix}\begin{pmatrix}-b_x s_x-b_y s_y-b_z s_z -i b_x s_y+i b_y s_x & -b_z s_x+b_x s_z+i b_z s_y-ib_y s_z\\b_z s_x-b_x s_z+i b_z s_y-ib_y s_z&-b_x s_x-b_y s_y-b_z s_z +i b_x s_y-i b_y s_x\end{pmatrix} \begin{pmatrix}0\\1
\end{pmatrix}\Bigg\}, \\
&=\frac{1}{2}((-s_x b_x -s_y b_y -s_z b_z -i s_x b_y +i s_y b_x)+(-s_x b_x -s_y b_y -s_z b_z +i s_x b_y -i s_y b_x)), \\
&= -s_x b_x - s_y b_y - s_z b_z = {\bf s_2}\cdot {\bf b}.
\end{align}
Since each detection station detects a single particle, it is discerned that the particle is in either the "up" or "down" state upon detection.

\section{Proof of Eq. (24)}
Let us begin by expanding the two expressions from eq.(22) in the standard form as follows with ${\bf r_a} = ({\bf a}\times {\bf s_1})$, ${\bf r_b} = ({\bf s_2}\times {\bf b})$ and ${\bf s_1} + {\bf s_2} = 0$ using geometric algebra \cite{Doran, Christian1, Christian3}
\begin{align}
{\mathbf a}\,\; {\mathbf s}_1&={\mathbf a}\cdot {\mathbf s}_1 + I_3( r_{ax}\,{\bf e_x} + r_{ay}\,{\bf e_y} + r_{az}\,{\bf e_z}), \label{9-nn} \\
{\mathbf s_2}\,\; {\mathbf b}&={\mathbf s}_2\cdot {\mathbf b} + I_3( r_{bx}\,{\bf e_x}  + r_{by}\,{\bf e_y} + r_{bz}\,{\bf e_z}). \label{8-nn}
\end{align}
The geometric algebra expression $\mathbf {GA}({\mathbf a},\, {\mathbf b},\,{\mathbf r}_{0})$ is then the product of these expressions and can be evaluated as follows:
\begin{align}
\mathbf {GA}({\mathbf a},\, {\mathbf b},\,{\mathbf r}_{0})&=\{{\mathbf a}\,\; {\mathbf s_1}\}\{{\mathbf s_2}\,\; {\mathbf b}\}, \label{115-n} \\
&=\{{\mathbf a}\cdot {\mathbf s}_1 + I_3{\mathbf r}_a\,\} \{{\mathbf s}_2\cdot {\mathbf b} + I_3{\mathbf r}_b\}, \\
&=({\mathbf a}\cdot {\mathbf s}_1)({\mathbf s}_2\cdot {\mathbf b}) + (I_3{\mathbf r}_b)({\mathbf a}\cdot {\mathbf s}_1) + (I_3{\mathbf r}_a)({\mathbf s}_2\cdot {\mathbf b}) + (I_3{\mathbf r}_a)(I_3{\mathbf r}_b), \\
&= ({\mathbf a}\cdot {\mathbf s}_1)({\mathbf s}_2\cdot {\mathbf b}) + (I_3{\mathbf r}_b)({\mathbf a}\cdot {\mathbf s}_1) + (I_3{\mathbf r}_a)({\mathbf s}_2\cdot {\mathbf b})-({\mathbf r}_a\cdot{\mathbf r}_b)-I_3({\mathbf r}_a\times{\mathbf r}_b),\\
&= ({\mathbf a}\cdot {\mathbf s}_1)({\mathbf s}_2\cdot {\mathbf b})-({\mathbf r}_a\cdot{\mathbf r}_b) + (I_3{\mathbf r}_b)({\mathbf a}\cdot {\mathbf s}_1) + (I_3{\mathbf r}_a)({\mathbf s}_2\cdot {\mathbf b})-I_3({\mathbf r}_a\times{\mathbf r}_b),\\
&= ({\mathbf a}\cdot {\mathbf s}_1)({\mathbf s}_2\cdot {\mathbf b})-({\mathbf a}\times {\mathbf s_1})\cdot({\mathbf s_2}\times {\mathbf b}) + (I_3{\mathbf r}_b)({\mathbf a}\cdot {\mathbf s}_1) + (I_3{\mathbf r}_a)({\mathbf s}_2\cdot {\mathbf b})-I_3({\mathbf r}_a\times{\mathbf r}_b),\\
&= ({\mathbf a}\cdot {\mathbf s}_1)({\mathbf s}_2\cdot {\mathbf b})-({\mathbf a}\cdot{\mathbf s}_2)({\mathbf s}_1\cdot{\mathbf b})+
({\mathbf a}\cdot{\mathbf b})({\mathbf s}_1\cdot{\mathbf s}_2) + (I_3{\mathbf r}_b)({\mathbf a}\cdot {\mathbf s}_1) + (I_3{\mathbf r}_a)({\mathbf s}_2\cdot {\mathbf b})-I_3({\mathbf r}_a\times{\mathbf r}_b),\\
&= -({\mathbf a}\cdot {\mathbf s}_2)({\mathbf s}_2\cdot {\mathbf b})+({\mathbf a}\cdot{\mathbf s}_2)({\mathbf s}_2\cdot{\mathbf b})-
({\mathbf a}\cdot{\mathbf b}) + (I_3{\mathbf r}_b)({\mathbf a}\cdot {\mathbf s}_1) + (I_3{\mathbf r}_a)({\mathbf s}_2\cdot {\mathbf b})-I_3({\mathbf r}_a\times{\mathbf r}_b),\\
&= -({\mathbf a}\cdot {\mathbf b}) + I_3\{{\mathbf r}_b({\mathbf a}\cdot {\mathbf s}_1) + {\mathbf r}_a({\mathbf s}_2\cdot {\mathbf b})-({\mathbf r}_a\times{\mathbf r}_b)\}, \\
&= -{\mathbf a}\cdot {\mathbf b} + {\bf r_0}.
\end{align}
For eq.(B6), we expanded the product of the last term in eq.(B5).

\end{document}